\documentclass[
 reprint,
superscriptaddress,
 amsmath,amssymb,
 aps,
]{revtex4-2}
\usepackage{xcolor}
\usepackage{graphicx}
\usepackage{dcolumn}
\usepackage{bm}
\usepackage{gensymb}
\usepackage{soul}
\usepackage{lineno}
\usepackage{amsmath}
\usepackage{braket}

\begin{document}

\preprint{APS/123-QED}

\title{Probing Interface-Driven Mechanisms of Non-Classical Light in van der Waals Heterostructures}

\author{Bárbara L. T. Rosa}\email{bltr@unicamp.br}
\affiliation{Institute for Physics and Astronomy, Technische Universität Berlin, 10623 Berlin, Germany}
\affiliation{Institute of Physics “Gleb Wataghin”, State University of Campinas, 13083-859 Campinas, Brazil}

\author{Lara Greten} 
\affiliation{Institute for Physics and Astronomy, Technische Universität Berlin, Berlin, Germany}

\author{Raphaela de Oliveira}
\affiliation{Brazilian Synchrotron Light Laboratory (LNLS), Brazilian Center for Research in Energy and Materials (CNPEM), Campinas, Sao Paulo 13083-970, Brazil}

\author{César Ribahi}
\affiliation{Department of Physics, Federal University of São Carlos, São Carlos, SP 13565-905, Brazil}

\author{Aris Koulas-Simos}
\affiliation{Institute for Physics and Astronomy, Technische Universität Berlin, 10623 Berlin, Germany}

\author{Chirag C. Palekar}
\affiliation{Institute for Physics and Astronomy, Technische Universität Berlin, 10623 Berlin, Germany}

\author{Yara Gobato}
\affiliation{Department of Physics, Federal University of São Carlos, São Carlos, SP 13565-905, Brazil}

\author{Ingrid D. Barcelos}
\affiliation{Brazilian Synchrotron Light Laboratory (LNLS), Brazilian Center for Research in Energy and Materials (CNPEM), Campinas, Sao Paulo 13083-970, Brazil}

\author{Andreas Knorr}
\affiliation{Institute for Physics and Astronomy, Technische Universität Berlin, 10623 Berlin, Germany}

\author{Stephan Reitzenstein}
\affiliation{Institute for Physics and Astronomy, Technische Universität Berlin, 10623 Berlin, Germany}






\begin{abstract}

Single-photon emitters in two-dimensional semiconductors offer a versatile platform for integrated quantum photonics, yet their performance is strongly influenced by local dielectric environments and substrate-induced disorder. Here, we examine SPEs in monolayer WSe$_2$ incorporated into hBN/WSe$_2$/Clinochlore van der Waals heterostructures and assess how interface-mediated dielectric modulation governs their optical and quantum characteristics. Low-temperature micro-photoluminescence reveals narrow emission lines (100 - 300 $\mu$eV) and robust non-classical behavior, with $g^{(2)}(0) = 0.13 \pm 0.02$ on SiO$_2$ and $0.54 \pm 0.02$ for emitters directly coupled to Clinochlore. Magneto-optical measurements yield effective g-factors near -8, consistent with defect states hybridized with dark excitons. WSe$_2$ on Clinochlore exhibits up to a fivefold enhancement in emission intensity, attributed to coupling with Fe-related substrate states that introduce resonant absorption near 1.75 eV. Kelvin probe force microscopy confirms strong dielectric contrast across thin and thick Clinochlore regions. Time-resolved photoluminescence shows that emitters on SiO$_2$ display a single $\approx 4$ ns lifetime, whereas those on Clinochlore exhibit biexponential dynamics with sub-nanosecond and tens-of-nanoseconds decay components. A phenomenological model incorporating coupling to bright and dark Fe-related states in Clinochlore accounts for modified excitation pathways. These results establish interface dielectric engineering in vdW heterostructures as an effective strategy for tailoring the radiative dynamics and brightness of quantum emitters in atomically thin materials.


\end{abstract}


\maketitle

\section*{Introduction}

Single-photon emitters (SPEs) have been described as a key element in quantum optical studies and technology. Significant efforts have been undertaken to develop systems that exhibit the characteristics of a high-quality non-classical light emitter: high brightness, strong multi-photon suppression, and high photon indistinguishability. While the focus of such developments has been mainly on traditional solid-state systems such as color centers in diamond\cite{Marshall2011,vanDam2019,Bradac2019,DOHERTY20131} and semiconductor quantum dots\cite{Heindel2023,GarcadeArquer2021}, more recently also research has been carried out to fabricate SPEs based on 2D materials that satisfy the mentioned conditions attributed to a quantum light source\cite{MichaelisdeVasconcellos2022,Esmann2024,Paralikis2025,Wu2025}. These systems are particularly attractive due to their exceptional versatility, offering broad tunability through strain engineering, control of the dielectric environment, and heterostructure design.

The first publications on transition metal dichalcogenide (TMDC) SPEs, specifically WSe$_2$ monolayers\cite{Kumar2015,Koperski2015,He2016,Kumar2016}, reported the intrinsic presence of quantum emitters, commonly arising from local disorders, such as edges\cite{Koperski2015} or wrinkles\cite{Iff2019}. Later, local defects and strain were engineered on the same material to achieve site-controlling and high-brightness single-photon emission\cite{Berraquero2016,Parto2021,von2023temperature,serati2024probing,Singh2025,Piccinini2025,Yu2025}. Moreover, although a clear explanation has yet not been established, the current description of the SPEs nature in WSe$_2$ systems relies on the presence of intrinsic defects in the gap (usually Se vacancies) combined with microscopic strained variation\cite{Linhart2019}; the local strain applied on the flake, for instance, via nanopillars or wrinkles lowers the dark exciton energies; thus, the system may experience the presence of hybridized states formed by defect states strongly localized in the band gap of the TMDC conduction band that behave as a radiative recombination channels. In contrast, other TMDCs such as WS$_2$\cite{Loh2024}, MoS$_2$\cite{Barthelmi2020}, and MoSe$_2$\cite{Yu2021} have also been presented in the literature as SPE candidates for quantum emission from radiative defects induced by nanoscopic ionic implantation \cite{Parto2021,Klein2019}. Additional classes of 2D semiconductor heterostructures formed by twisted staked TMDC monolayers have also emerged as promising candidates for single-photon emission. In those systems, the key novelty arises when twisted bilayers create a moiré superlattice, which localizes IXs in the resulting periodic potential\cite{Andrei2021,RuizTijerina2020,Frg2021}. This confinement leads to discrete states that can act as sources of non-classical light\cite{Seyler2019,Baek2020}. Further, it was observed that 2D materials beyond TMDCs can behave as SPEs, depending on the processing steps they underwent. Graphene-related structures\cite{Zhao2018} or local defects on hexagonal-Boron Nitride (hBN)\cite{Grosso2017,Zeng2024,Xu2018,Stern2024} are  systems where SPEs have been observed even at room temperature; the latter, an insulator, showing striking properties, such as quantum coherence response similar to those observed in color centers in diamond\cite{Marshall2011,vanDam2019}. In parallel with the active material exploration, substantial fabrication efforts, such as integration with circular Bragg grating resonators\cite{Iff2021} or Fabry–Pérot cavities\cite{Flatten2018}, have been carried out to achieve an optical response comparable to that of well-established SPEs. This quantum performance limitation may be intrinsically linked to the pronounced environmental sensitivity of 2D material-based SPEs, especially to proximity effects and dielectric disorder, mechanisms that have not yet been systematically investigated.

In this work, we explore the role of the dielectric environment in the optical and quantum topical properties of SPEs generated in van der Waals (vdW) heterostructures composed of WSe$_2$ and the layered phyllosilicate material Clinochlore, a dielectric vdW mineral that carriers outstanding surface properties and high density of impurities\cite{deOliveira2022,deOliveira2024,Kawahala2025}. 
By performing magneto-optical  measurements, we demonstrate the existence of localized emitters and their origin from intrinsic defect states energetically located near the bottom of the conduction band, hybridized with dark exciton states. Additionally, our experimental results demonstrate a pronounced interface coupling between the layered materials, as evidenced by a photoluminescence (PL) quantum yield enhancement up to 5 times, that is attributed to intrinsic iron (F) impurities in the Clinochlore lattice \cite{deOliveira2022}. Kelvin probe force microscopy (KPFM) reveals an evident dielectric contrast dependence with the presence of Clinochlore layer and its thickness, which attests that the optical changes are related variations in the dielectric environment of the heterostructure. Results from time-resolved photoluminescence (TRPL) measurements confirms the Coulomb interaction landscape by drastically altering the carrier decay dynamics.

To support our experimental observations, we introduce a phenomenological model that describes coupling mechanisms in the van der Waals heterostructure. Finally, second-order autocorrelation measurements reveal the effects of interface coupling on the quantum nature of the emitters. Auto-correlation values of $\text{g}^{(2)}(0) =0.13\pm0.02$ and $0.54\pm0.02$ extracted from regions with SiO$_2$ and Clinochlore, respectively, reflects the environment plays an active and previously underappreciated role in the operation of two-dimensional single-photon emitters, and must therefore be treated as a design parameter rather than a secondary consideration. Our findings provide direct experimental evidence that unveils the role of interface-induced dielectric modulation as a key parameter for engineering and controlling high-quality SPEs in van der Waals heterostructures.

\section*{Experimental Results}

\begin{figure*}[!htb]
    \centering
    \includegraphics[width=0.8\textwidth]{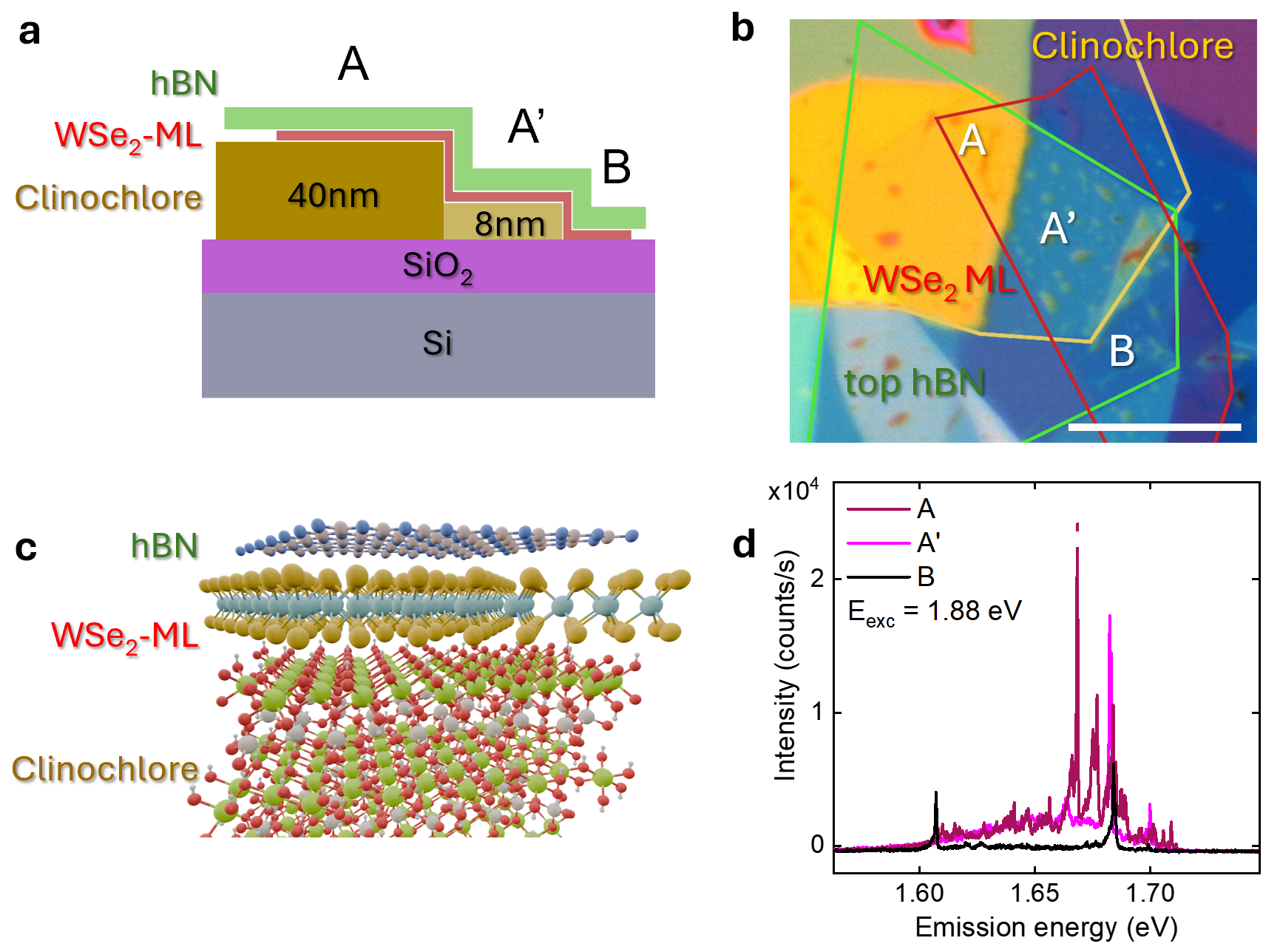}
    \caption{\textbf{(a)} Schematic view of a van der Waals heterostructure stacked on SiO$_2$/Si substrate. \textbf{(b)} Optical image of the architecture illustrated in (a). \textbf{(c)} Lamellar structure of hBN, WSe$_2$ and Clinochlore, respectively. \textbf{(d)} $\mu$PL response at 4~K extracted from regions A (Thick-Clinochlore substrate), A' (Thin-Clinochlore substrate), and B (SiO$_2$ substrate), as indicated in (a). The PL intensity increases up to 5 times depending on the Clinochlore thickness compared with the bare SiO$_2$ substrate.}
    \label{fig1}
\end{figure*}

We fabricated heterostructures composed by hBN/WSe$_2$/Clinochlore using mechanical exfoliation and dry-transfer method techniques \cite{Kim2016}, as described in the Methods section. A schematic picture of the device is shown in Figure~\ref{fig1}a while Figure~\ref{fig1}b displays an optical image of our heterostructure. In this work, we studied three distinct regions, hBN/WSe$_2$/thick-Clinochlore/SiO$_2$ (region A), hBN/WSe$_2$/thin-Clinochlore/SiO$_2$ (region A') and hBN/WSe$_2$/SiO$_2$ (region B), indicated in Figure~\ref{fig1}a, corresponding to different substrate layers. Clinochlore is particularly suited for this study because it simultaneously provides a clean van der Waals interface, tunable dielectric screening through thickness control, and intrinsic doping that introduces optically active internal states, enabling the controlled investigation of interface-mediated mechanisms.

Considering that the properties of different exfoliated flakes may change from sample to sample, we deposited the same WSe$_2$ ML over the three regions to be investigated. Therefore, A, A' and B characterize, respectively, the WSe$_2$ in direct contact with a thick Clinochlore region ($\sim40$~nm), a thin Clinochlore region ($\sim8$~nm) and SiO$_2$ ($285$~nm)/Si substrate. The sample is covered by a thin ($5$~nm) hBN flake (Fig.~\ref{fig1}a). Figure~\ref{fig1}c shows the crystalline structure of hBN, WSe$_2$ and Clinochlore, in which we see that, although all crystals are composed of lamellar atomic structures, the Clinochlore presents a more complex crystallographic composition\cite{deOliveira2022,deOliveira2024}. Here, the charge-neutral T-Oc-T stacking is the basic structure of trioctahedral 2:1 phyllosilicates, in which the Oc layer is formed purely by $\text{Mg}^{2+}$ ions, such as talc. The Clinochlore structure is a modification of talc, in which one-fourth $\text{Si}^{4+}$ ions in each T-layer is substituted by an $\text{Al}^{3+}$ ion. Intercalating the T-Oc-T stacking, a further octahedral layer of $\text{Mg}^{2+}$/$\text{Al}^{3+}$ ions with O/OH at the vertices (the brucite-like hydroxide layer). 

To study the optimal response of the sample it was excited using a laser energy of E$_{\text{exc}}=1.88~\text{eV}$, above the WSe$_2$ neutral exciton emission ($\text{X}^0\sim1.74~\text{eV}$)\cite{Lyons2019}. The resulting PL signal obtained at $4\text{K}$ from A, A' and B are displayed in Figures~\ref{fig1}d,e.  Most prominently, we observe the appearance of sharp lines near and below the excitonic emission of the WSe$_2$ monolayer
, with an average linewidth between 100 and 300~$\mu\text{eV}$, which indicates the presence of SPE candidates, as widely discussed in the literature of quantum emitters based on semiconducting 2D materials \cite{RobertodeToledo2025,Berraquero2016,Parto2021,von2023temperature,serati2024probing,Singh2025,Piccinini2025,Yu2025}. Additionally, we mention that similar to what occurred in previous reports, the appearance of SPEs in WSe$_2$ mono- or bilayers is a process that depends mainly on the lattice defects (e.g.~intrinsic vacancies) and local strain, commonly induced by elements on the substrate (roughness\cite{serati2024probing}, wrinkles\cite{Iff2019}, nanopillars\cite{Berraquero2016,Parto2021,Sortino2021}, etc.), not requesting any further direct processing on the flake to be generated. Moreover, a remarkable up to 5-fold PL enhancement from the B region is observed in Figure~\ref{fig1}d, whereas regions A and A' show similar PL intensities but sharp lines emission located at different energies (see SI, Sec.1, which exhibits several additional spectra, confirming the overall behavior from regions A, A' and B).

\subsection*{Optical and magneto-optical properties of SPEs}

To concentrate on the novelties mainly derived from the Clinochlore as a new substrate, the results displayed from this point are obtained from the data extracted at region A, consisting of WSe$_2$/thick-Clinochlore, unless further comparison between regions is required to enrich our discussion. 

\begin{figure*}[!htb]
    \centering
\includegraphics[width=0.8\textwidth]{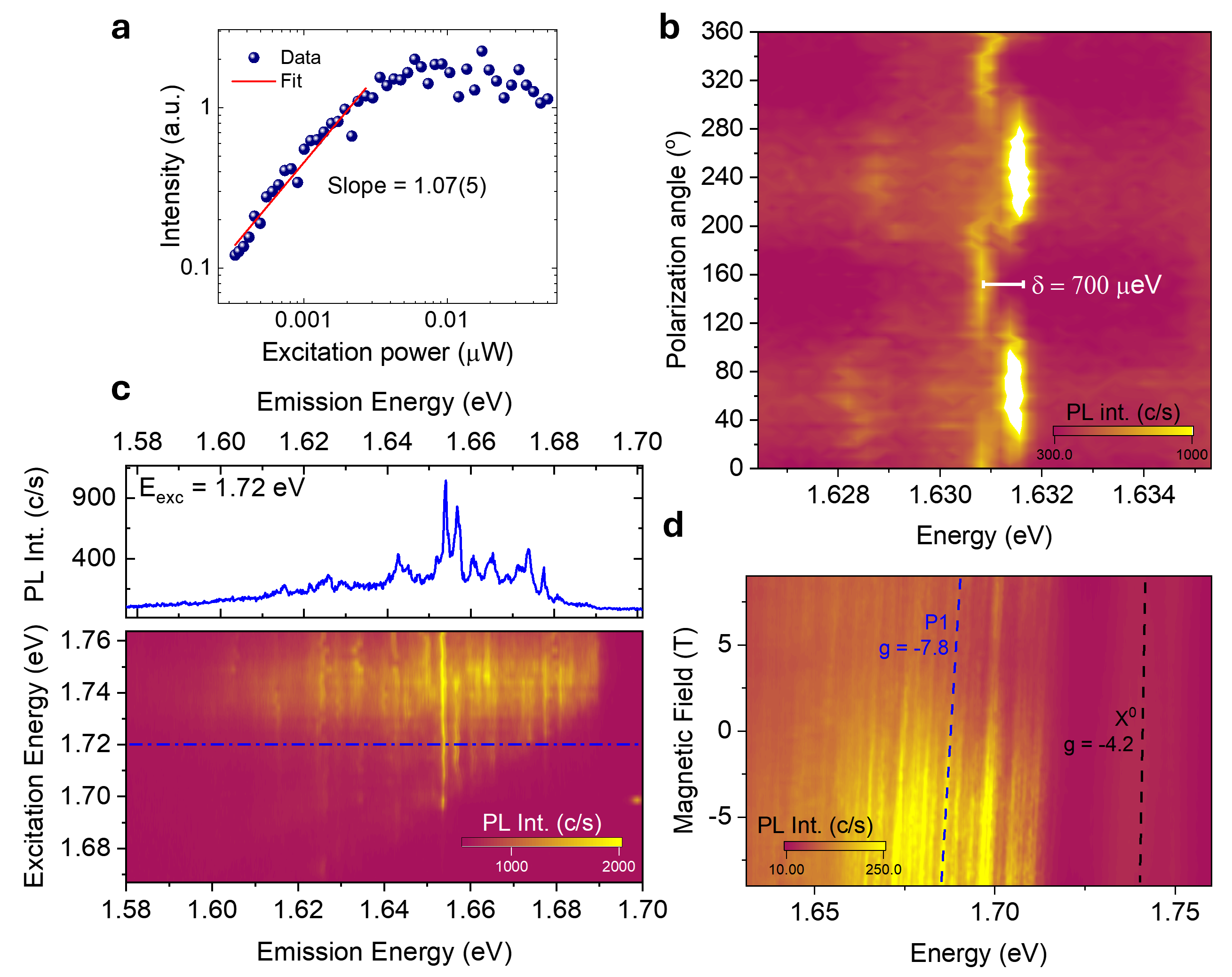}
    \caption{$\mu$PL and magneto-PL measurements. Results of \textbf({a}) power- dependent and \textbf({b}) polarization-dependent characterization revealing an FSS of $\sim700~\mu\text{eV}$. \textbf({c}) PLE studies, in which the spectrum shown in the upper panel was extracted under excitation energy of 1.72 eV (see dashed line in the lower part) while the color-map shows the emission response of several SPEs under an excitation energy range from 1.65 to 1.77 eV. \textbf({d}) Color-coded map of the circularly polarized PL intensity of WSe$_2$/Clinochlore as a function of magnetic field. The dashed lines indicate the PL peak energy as a function of the magnetic field. The circularly polarized PL spectra ($\sigma^-$ for positive magnetic fields) were recorded at 3.6 K using a linearly polarized laser excitation with an energy of 1.88 eV PL spectra of WSe$_2$/Clinochlore for selected perpendicular magnetic fields. The labels P1 and X$^{0}$ indicate the peak emission of defect-states bound to an exciton and the neutral bright exciton peak, respectively.}
    \label{fig2}
\end{figure*}

Among the characteristics expected to be observed in a SPE candidate, Figure~\ref{fig2}a shows the excitation power dependence of PL intensity for an emission line at 1.65~eV in double logarithmic scaling. The fit to the data that yields a linear slope of 1.07(5) and a PL intensity saturation for excitation power values exceeding 4 $\mu$W. Additionally, as shown in Figure~\ref{fig2}b, a degree of linear polarization of $82\%(6)$ was determined via polarization-resolved $\mu$PL measurements. The data were collected from region A, but similar power and polarization-dependent properties were found for numerous SPEs also arising from A' and B regions. The polarization-resolved $\mu$PL of an emitter under non-resonant excitation also shows the presence of pairs of correlated emission lines and a fine structure splitting (FSS) $\delta$ $\sim700~\mu\text{eV}$ at zero-field, independent of the sample region. 

To achieve further insights into the effects of Clinochlore on the SPE properties, we performed photoluminescence excitation (PLE) spectroscopy over an extended spectral range. Here, the excitation laser was scanned from 705~nm to 756~nm (1.76~eV to 1.64~eV) with a step of 1~nm and laser straylight was suppressed in detection using a tunable long-pass optical filter. The results are depicted in Figure~\ref{fig2}c as a PLE intensity color-code map. The blue curve depicts a spectrum at E$_{\text{exc}}=1.72~\text{eV}$ as an example of the spectra that compose the PLE map. Although we observe multiple resonances, the strongest response arises from the energy excitation around the excitonic complexes emission at 1.70~eV, contrary to previous reports\cite{Tonndorf2015,vonHelversen_2023}, in which a $\mu$PL intensity increases with decreasing the energy difference between the excitation and the emission lines was reported. Nonetheless, studies have also shown that the dominant PL intensity response may come from excitation energies around the excitonic emissions \cite{serati2024probing,Sortino2021}, therefore bringing out an understanding that the recombination process of those emitters might be additionally associated with the environment, Clinochlore in our case,  where the material is located.

In order to obtain a deeper understanding of the excitonic nature of the SPE lines, magneto-optical spectroscopy was conducted, which reveals information about g-factors. Figure~\ref{fig2}d shows the color-coded map of the circularly polarized PL intensity from the defect-mediated radiative recombination in our sample as a function of the magnetic field collected from the region A with WSe$_2$/thick-Clinochlore. The dashed lines present the magnetic field dependence of these sharp emission peaks. In the presence of an out-of-plane magnetic field (Faraday configuration), the  sharp emission peak energy shifts as
\begin{equation}
\text{E}_{i}\text{(B)} =\text{E}_{i}(0) \pm \frac{1}{2}\text{g}_{i}\mu_\text{B} \text{B},
\label{eq.magneticshifts}
\end{equation}
where $\mu_\text{B}$ is the Bohr magneton, $\text{B}$ the magnetic field strength, $i$ = 1,2,3 ... The $\pm$ sign depends on the valley degree of freedom, associated with the circular polarization $\sigma_{\pm}$, and the linear term corresponds to the Zeeman shift, with an effective Landé $g$-factor $g_{i}$. The $g$-factor values are directly related to the angular momenta of the valence and conduction band states involved in the exciton transition.

\begin{figure*}[!htb]
    \centering
    \includegraphics[width=0.8\textwidth]{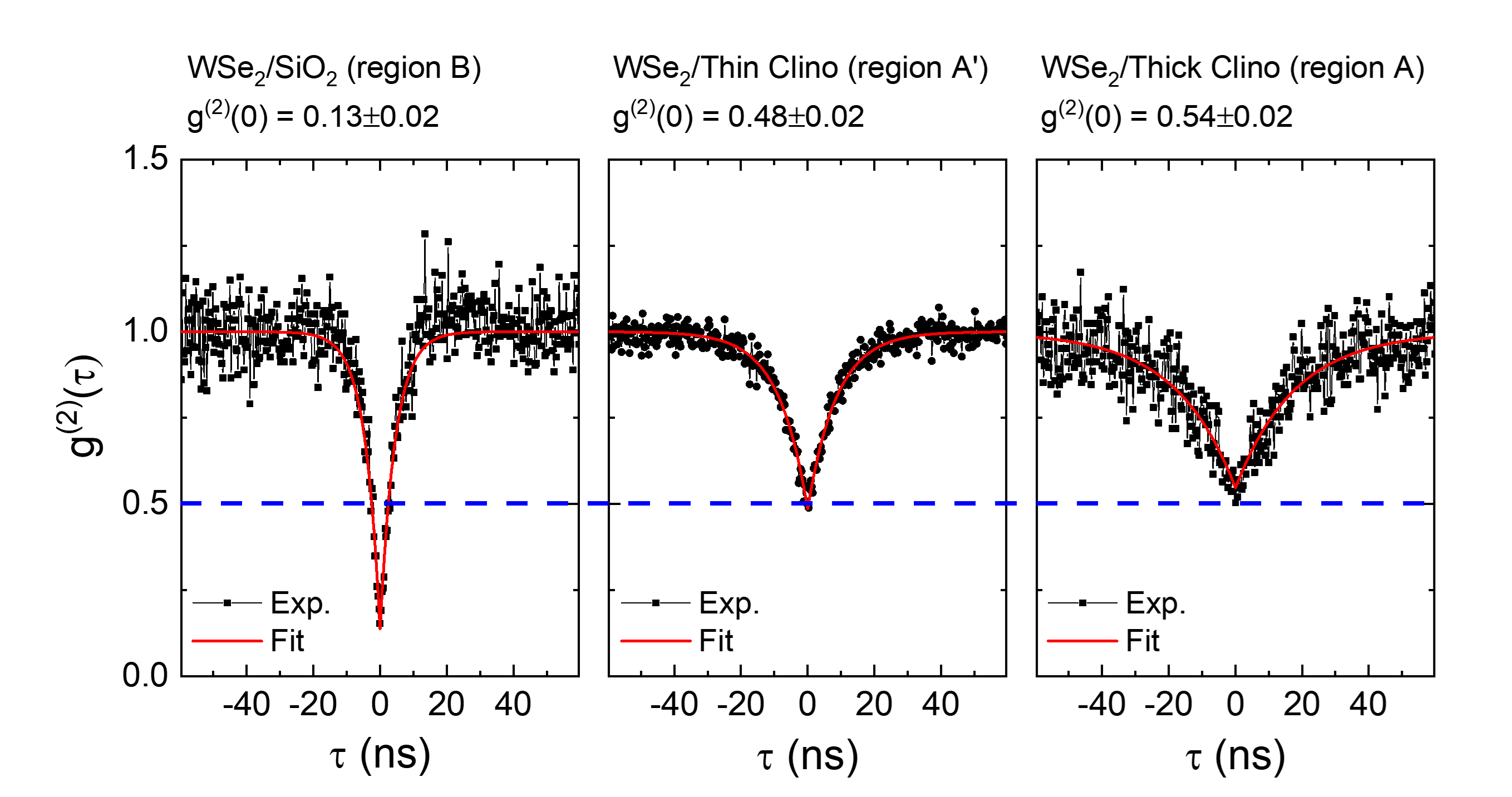}
    \caption{Second-order autocorrelation function under CW excitation revealing a photon antibunching value of $g^{(2)}(0) = 0.13 \pm 0.02$, $0.48 \pm 0.02$ and $0.54 \pm 0.02$, for substrates of SiO$_2$, thin Clinochlore, and thick Clinochlore, respectively. The horizontal dashed line indicates the SPE classical limit $g^{(2)}(0) = 0.5$.}
    \label{g2}
\end{figure*}

Based on Equation~\ref{eq.magneticshifts} and the data shown in Figure~\ref{fig2}, we extracted the $g$-factors for both the exciton and the sharp emission peaks. The determined $g$-factors of neutral bright excitons and trions have $g$-factor values of $g \approx -4.2$, which is in agreement with previous experiments \cite{Arora2018,Molas2019,Liu2019} and theoretical predictions \cite{FariaJunior-gfactors-2022} in the literature. The sharp emission peaks, on the contrary, have $g$-factor values of around -8, which are similar to the values reported for the dark exciton and trion lines in previous works \cite{Li2019,Forste2020,serati2024probing,Molas2019,Liu2019,He2020,Robert2017,Cavalini2024,SeratideBrito(2022)} as well as SPE-like features arising in WSe$_2$ flakes\cite{Dang2020} on different types of substrates. Notably, the magnetic field dependence observed for these sharp peaks suggests that its spin-valley configuration is similar to the configuration for non-localized dark exciton and neutral bright exciton in TMDCs\cite{Li2019a}. Particularly, the  presence of local strain usually localizes dark excitons and allows their hybridization with defect levels which create a new electron–hole pair configuration known as an intervalley defect bright exciton, resulting in an efficient radiative decay\cite{Linhart2019}. Therefore, our magneto-optical experiments reveal that thick-Clinochlore underneath likely does not influence the nature of SPEs in WSe$_2$. These findings are, therefore, consistent with the interpretation of the hybridization between localized defects and dark states of WSe$_{2}$. 

To validate the quantum nature of our emitters, we investigated the localized emission by Hanbury Brown and Twiss (HBT) measurements of the second-order autocorrelation function $\text{g}^{(2)}(\tau)$ (see Methods). Here, the multi-photon suppression is expressed by $\text{g}^{(2)}(0)$, which is highly sensitive to the presence of additional radiative channels and uncorrelated background emission, providing a direct and reliable probe of the emitter’s non-classical character. Figure~\ref{g2} shows the auto-correlation histogram obtained from regions A, A' and B under CW excitation at 660~nm. The quantum nature of emission is seen through the reduced coincidences at zero time delay. For a single-photon emitter, we expect $\text{g}^{(2)}(0) < 0.5$  (dashed blue line). The multi-photon suppression was quantized by fitting the data with a convolution of the theoretical solution of a two-level system (see SI for more details). The extracted values for regions B, A' and A are $\text{g}^{(2)}(0) = 0.13\pm0.02$, $0.48\pm0.02$ and $0.54\pm0.02$, respectively. The single-photon emission is clearly confirmed for the emitter in the WSe$_2$/SiO$_2$ region, in good agreement with the literature \cite{Kumar2015,vonHelversen_2023,Wu2025}. Multi-photon suppression close to the threshold of $\text{g}^{(2)}(0) = 0.5$ for the WSe$_2$/Clinochlore-generated emitters, indicates the presence of additional excitation and emission channels reducing the single-photon purity and a possible dependence of this effect on environment and layer thickness.

\subsection*{Influence of dielectric screening on the TMDC monolayer properties}


To investigate the consequences of the environment on the WSe$_2$ monolayer, one must consider that changes in the dielectric screening have significant influence on the Coulomb interactions between charge carriers in the case of a TMDC monolayer\cite{Raja2019,Adeniran2023,Raja2017,Noori2019}, and considerably impact its optoelectronic properties. So far, our findings have raised an enlightening discussion about the influence of a phyllosilicate substrate compared to a bare SiO$_2$/Si substrate and their effects. However, optical spectroscopy techniques used to date are not suitable to precisely investigate the 2D materials' interfaces, which requires more advanced scanning probes techniques to deepen into the differences between substrates. 

\begin{figure*}[!htb]
    \centering
    \includegraphics[width=1\textwidth]{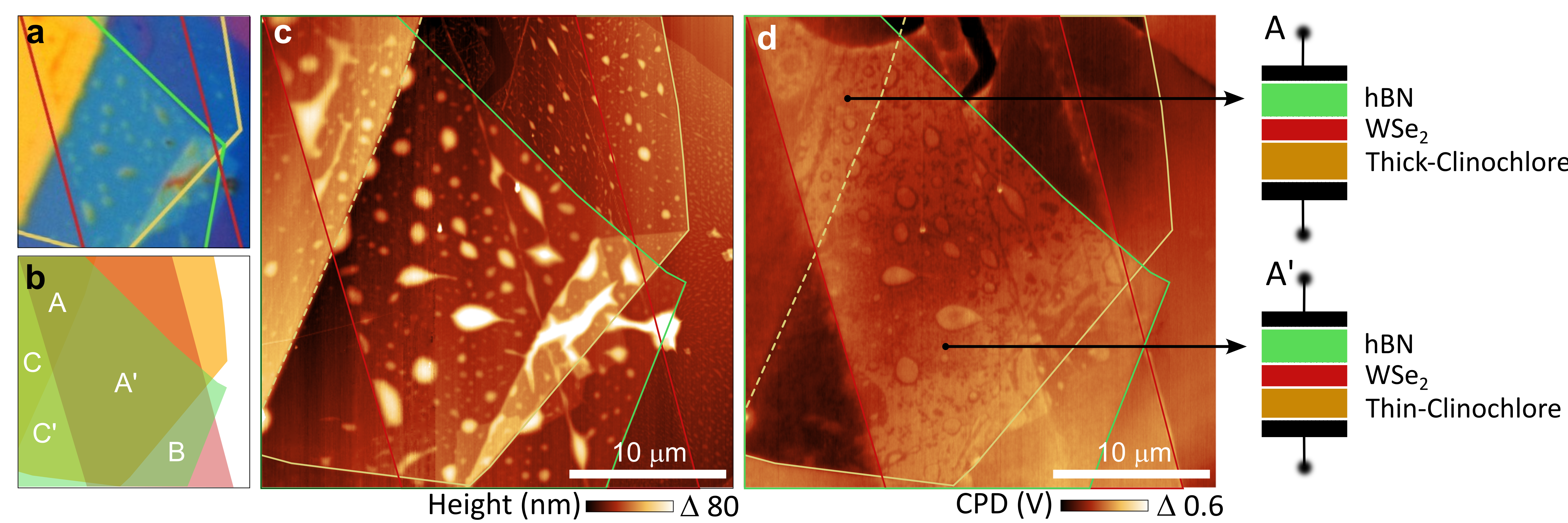}
    \caption{Dielectric contrast of the hBN/WSe$_2$/Clinochlore heterostructure probed by KPFM. \textbf{(a)} Optical microscopy image of a selected region from Fig.~\ref{fig1}a, with its corresponding \textbf{(b)} schematic of the stacked bottom Clinochlore (light yellow shade for the thin region and dark yellow shade for the thick region), WSe$_2$ (red shade), and top hBN (green shade). The A and A' regions correspond to heterostructure regions formed by hBN/WSe$_2$/thick-Clinochlore and hBN/WSe$_2$/thin-Clinochlore, respectively, while the B and B' regions correspond to hBN/thick-Clinochlore and hBN/thin-Clinochlore. The local capacitor formed by the stacking of the distinct dielectrics is depicted on the right side for regions A and A'. \textbf{(c)} Topography and \textbf{(d)} CPD images of the selected heterostructure region acquired simultaneously by single-pass scan KPFM at $<2\%$ relative humidity.}
    \label{fig3}
\end{figure*}

To further investigate the interface effects, we performed single-pass KPFM measurements on our sample. The optical microscopy image in Figure~\ref{fig3}a features both the hBN/WSe$_2$/thick-Clinochlore and hBN/WSe$_2$/thin-Clinochlore regions to determine whether the difference in Clinochlore thickness also implies a difference in its dielectric behavior, and thus, shed light onto the interface disorder derived from the substrate.  The KPFM technique operates by applying a tip-sample potential difference that nullifies the local contact potential difference (CPD) between them, providing both the topography and CPD map of the scanned region\cite{Sadeghi2012}. In this context, the system can be modeled as a capacitor, where the plates are the tip and the grounded substrate, while the dielectric medium is the hBN/WSe$_2$/Clinochlore heterostructure itself\cite{Sadeghi2012}. The total capacitance of a capacitor formed by stacking different dielectrics is a series association of individual capacitors formed by each dielectric. Thus, for an insulating heterostructure system, the main contrast in a CPD image acquired by KPFM arises from regions formed by the stacking of dielectrics with different dielectric constants $\varepsilon$.

The results of KPFM measurements are depicted in Figure~\ref{fig3}a, in which we see an optical image of the studied region selected, with its schematic dielectric stacking displayed in Figure~\ref{fig3}b. Additionally, the topography and CPD images acquired by KPFM are shown in Figure~\ref{fig3}c and d, respectively. We extracted a thickness of approximately 3.8(6) nm from the topography image for the thin Clinochlore layer, with a step of 41(3) nm to the thick region. It is possible to observe a clear CPD contrast between regions formed by hBN/thick-Clinochlore(C) and hBN/thin-Clinochlore(C’), as well as between regions formed by hBN/WSe$_2$/thick-Clinochlore(A) and hBN/WSe$_2$/thin-Clinochlore(A’). The C/A regions are lighter than the C’/A' regions, indicating that the Clinochlore exhibits distinct dielectric behavior depending on its number of layers. Light regions correspond to higher applied potential differences, whereas dark regions relate to lower applied potential differences. This result suggests that the light regions have a smaller capacitance than the dark regions and, consequently, a smaller effective dielectric constant. Therefore, the dielectric screening experienced by WSe$_2$ can vary depending on the Clinochlore thickness. We highlight that, despite the sample presenting several bubbles across the surface derived from monolayer dry-transfer process employed, the thin and thick Clinochlore regions show similar density and lateral size of objects. Moreover, the effects in the CPD contrast is barely noticed, whereas the emission yield trend is observed on the overall regions (as seen in Fig. S1, where several representative spectra are displayed), indicating that those experimental results are governed primarily by the local dielectric environment rather than by transfer-induced structural disorder.



It is worth mentioning that Clinochlore is a naturally hydrated mineral with a negatively charged silicon oxide tetrahedral layer\cite{deOliveira2022} that would be exposed once the material is exfoliated\cite{deOliveira2022}. Therefore, based on this information combined with our experimental findings, we propose two explanations for the observed differences in dielectric behavior depending on Clinochlore's thickness: \textit{(i)} the surface potential of the negatively charged layer overrules in the case of thin-Clinochlore, i.e., that the thinner is the region, the lower is the CPD response; or \textit{(ii)} considering the results reported in ref.\cite{deOliveira2024}, the thinner the Clinochlore flake is, the less hydrated it appears, and consequently, the lower is the CPD value. Furthermore, since the sample was annealed during the fabrication process (see Methods), the thin Clinochlore region may have become drier than the thicker region, leading to the observed dielectric contrast.

\subsection*{Description of interface-mediated decay processes in TMDC SPEs}


Although we have provided several single-emitter signatures from a WSe$_2$ ML, we also carefully navigate our results to provide additional insights into the nature of those intrinsic SPEs. To unravel the enhanced quantum yield (QY) and the substrate dependent time decay changes, we combine our experimental and theoretical findings to detach the response associated with the presence of a phyllosilicate substrate underneath WSe$_2$ ML from the effects of Clinochlore thickness variation. To date, the current understanding relies on the combination of localized defect states, strain and the existence of dark states, for which, as Linhart et al.\cite{Linhart2019} first discussed, when the system contains all those elements, a hybridization of dark exciton energy levels and intra-gap intrinsic localized defects occurs due to the strain bending the TMD conduction band levels into the defect states. However, the mechanism described so far does not account for the role of the interface WSe$_2$/substrate. This omission may contribute to the discrepancies observed across experimental reports employing nominally similar device architectures, suggesting that interface effects could play a decisive role in the unintentional appearance of SPEs in WSe monolayers.


\begin{figure*}[!htb]
    \centering
    \includegraphics[width=0.9\textwidth]{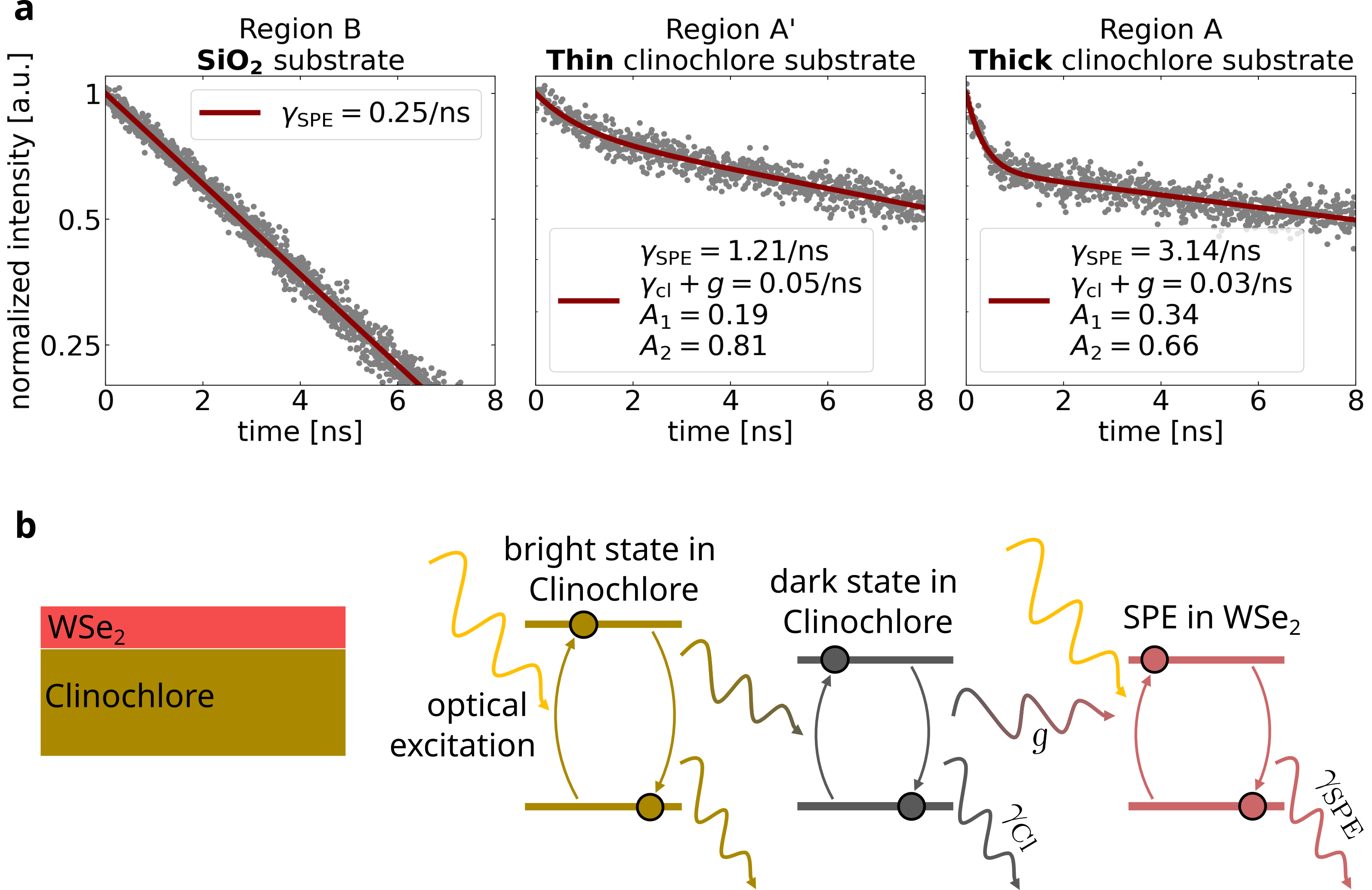}
    \caption{ TRPL experimental results and decay model of our quantum emitters. (\textbf{a}) For region B a mono-exponential decay $e^{-\gamma_\mathrm{SPE}t}$ is observed. For regions A' and A, a biexponential decay is fitted using Eq.~\eqref{eq: solution for SPE decay}. Uncertainties for the fitting parameters are given in the SI. The data is normalized such that $A_1+A_2 = 1$. (\textbf{b}) Sketch of the proposed emission process: optical excitation populates a bright Fe-related state in Clinochlore, which relaxes non-radiatively into an energetically lower dark Clinochlore state. This dark state couples to the SPE in WSe\textsubscript{2} with a rate $g$, while the SPE decays with $\gamma_\mathrm{SPE}$ and the dark Clinochlore state with $\gamma_\mathrm{Cl}$.}
    \label{trpl}
\end{figure*}


Beyond the established picture of WSe$_2$-based SPE formation, multiple radiative recombination channels may also contribute to the observed emission. In this context, the effective g-factor extracted from magneto-PL measurements serves as an important indicator of the underlying radiative pathway \cite{Dang2020}. In our case, the predominant g-factor of $\sim9$ suggests that the emission likely originates from an intra-gap defect state located near the conduction band, which recombines radiatively with the valence band.
Although the description based on magneto-response may shed light on the nature of TMDC-SPEs, a clear picture of the decay dynamics, which ultimately impact the lifetime, brightness, spectral stability, and temporal purity of the non-classical emission, is still elusive and requires careful investigation. To this end, we performed time-resolved photoluminescence (TRPL) measurements on the three target regions described above. Considering that proximity effects can influence both the intrinsic properties of defect states (energy, symmetry, and radiative pathways) and their extrinsic stability (spectral noise, quenching, and time decay), the distinct interaction regimes between the WSe$_2$ and the substrate, our combined (TR)PL and KPFM measurements can provide key insights into the decay mechanisms governing WSe$_2$-based SPEs and, consequently, the overall optical performance of the emitter.

In the TRPL data shown in Figure \ref{trpl}a, we first notice a substantial difference in lifetime between regions with different substrates beneath WSe$_2$ (SiO$_2$ or thin and thick Clinochlore layers). See Methods for measurements information. While the emitters arisen from region B follow a simple exponential decay $\gamma_\mathrm{SPE}=0.25\,\text{ns}^{-1}$, the decay extracted from A and A' regions clearly exhibit two distinct timescales. To explain the behavior originating from the Clinochlore interface, it is worth depicting the electronic properties of the layered material. Previous work has shown that Clinochlore exhibits optical absorption associated with iron (Fe) impurities. In particular, Ref.~\cite{deOliveira2022} reports a Fe$^{2+} \rightarrow$ Fe$^{3+}$ transition with an absorption peak around 1.75~eV, which coincides with the quasi-resonant excitation energy used to pump the SPE. This scenario indicates that, in contrast to SiO$_2$, the Clinochlore substrate is optically active and absorbs energy within the relevant excitation range. While similar materials typically re-emit absorbed energy through fluorescence, no detectable fluorescence signal is observed from Clinochlore under comparable conditions. 

We therefore hypothesize that Clinochlore hosts energetically lower-lying, dipole-forbidden dark states that serve as efficient non-radiative relaxation pathways for the bright Fe-related excited states. This mechanism naturally accounts for the absence of fluorescence and suggests the existence of a long-lived energy reservoir that can influence the decay dynamics of the SPE.
To exploit this scenario for a qualitative explanation of the unusual two-timescale decay of the SPE emission on Clinochlore substrates, we introduce a minimal model, illustrated in Fig.~\ref{trpl}b. The model comprises the SPE state, a bright Clinochlore state, and a dark Clinochlore state:
The laser populates both the SPE and the bright Clinochlore state. The latter is assumed to relax rapidly into the dark Clinochlore state, on a timescale comparable to the initial population of the SPE. In contrast, the dark Clinochlore state exhibits a longer lifetime enabling an excitation transfer to the SPE. This delayed feeding of the SPE gives rise to the second, slower decay component, TRPL$\sim \sigma_\text{SPE}(t)$, following the SPE occupation dynamics \cite{kira_quantum_1999}.
For a quantitative analysis, we reduce the model to the occupations of the SPE and the dark Clinochlore state, $\sigma_\mathrm{SPE}$ and $\sigma_\mathrm{Cl}$, and assume that both states are initially populated. The corresponding coupled rate equations read
\begin{align}
\partial_t \sigma_{\mathrm{SPE}} &=
-\,{\gamma_{\mathrm{SPE}}\,\sigma_{\mathrm{SPE}}}
+\,{g\,\sigma_{\mathrm{Cl}}} \label{eq: differential equation for SPE 1}, \\
{\partial_t\,\sigma_{\mathrm{Cl}}} &=
-\,{\gamma_{\mathrm{Cl}}\,\sigma_{\mathrm{Cl}}}
-\,{g\,\sigma_{\mathrm{Cl}}}, \label{eq: differential equation for dark Clino state}
\end{align}
which are solved by
\begin{align}
 {\sigma_{\mathrm{SPE}}(t)} &
 {\,= {{A_1}}  e^{-\gamma_{\mathrm{SPE}} t} + {A_2}  e^{-(\gamma_{\mathrm{Cl}} + g) t}}. \label{eq: solution for SPE decay}
\end{align}
The weights $A_{1/2}$ of the two decay components can be expressed in terms of the initial occupations as well as the decay and coupling rates (see SI for details). Equation~\eqref{eq: solution for SPE decay} is used to fit the TRPL in Figure~\ref{trpl}a, normalized under the constraint $A_1 + A_2 = 1$.
The decay rates extracted from the fits are summarized for the different substrates in Table~\ref{table:interpretation_of_decay_rates}.

\begin{table}[h]
\centering
\caption{Decay rate trends for different substrates}
\label{table:interpretation_of_decay_rates}

\setlength{\tabcolsep}{10pt} 
\renewcommand{\arraystretch}{1.15} 

\begin{tabular}{c|c c c|c}

      & B   & A'        & A         & trend \\
\hline\hline
$n$   & 1.5 & 2--2.1$^{\mathrm{a}}$ & 2--2.1$^{\mathrm{b}}$ & $\uparrow$ \\
\hline
$\gamma_{\mathrm{SPE}}$ [ns$^{-1}$] & 0.25 & 1.21 & 3.14 & $\uparrow$ \\
\hline
$\gamma_{\mathrm{Cl}} + g$ [ns$^{-1}$] & -- & 0.05 & 0.03 & $\downarrow$ \\
\end{tabular}

\vspace{0.2em}
\footnotesize
Tendency to $^{\mathrm{a}}$lower / $^{\mathrm{b}}$higher values, see KPFM (Fig.~\ref{fig3})
\end{table}

The SPE decay rate $\gamma_\mathrm{SPE}$ increases with the substrate refractive index $n$, in qualitative agreement with Ref.~\cite{thranhardt_relation_2002}.

The slow decay component is governed by two contributions: the coupling rate $g$, and $\gamma_{\mathrm{Cl}}$ which accounts for other loss channels of the dark Clinochlore state.
The observed decrease of $\gamma_\mathrm{Cl}+g$ with increasing Clinochlore thickness can therefore be associated with two possible trends: (1) dark states becoming more stable, and/or (2) a reduced coupling efficiency to the SPE.
A similar discussion of the amplitudes $A_{1/2}$ is provided in the SI. Taking the different rates for the thick and thin Clinochlore substrates into account, the fitted values for $A_{1/2}$ are consistent with a significantly larger initial energy reservoir $\sigma_{\mathrm{Clino,dark}}\vert_{t=0}$ in the thick Clinochlore substrate.

As initially discussed, the sample was intentionally fabricated to contain three regions with different underneath substrates, and a single WSe$_2$ monolayer placed across all regions A, A' and B (shown in Fig.~\ref{fig1}). Such design was chosen to minimize disorders induced by mechanical exfoliation and dry-transfer process afterward, and, therefore, improve the reliability of the observed interface-driven mechanisms behind optical and quantum properties of those localized emitters. The mechanism proposed in Figure~\ref{trpl}b provides valuable insights into the decoherence channels observed in 2D-SPEs, specially WSe$_2$-based systems. Several works have also reported either double-exponential or single long decay components (several nanoseconds), for which a clear physical understanding remained lacking. As a consequence, the fabrication of high-quality single-photon emitters continues to be challenged by interface disorder, regardless of the substrate employed. 

The anti-bunching results further supports to our phenomenological interpretation. In the case of quantum dots, for example, brightness enhancement is typically accompanied by stronger multi-photon suppression, as it predominantly originates from a reduction of non-radiative recombination pathways. The WS$e_2$-based SPEs, on the contrary, may exhibit a non-intuitive behavior, which can be understood by considering the microscopic origin of these emitters. Because these defect states are expected to originate from the same type of vacancy, the resulting emitters can exhibit similar optical properties from site to site. The PL enhancement in such a system can be attributed to simultaneous occupation of energetically similar defect states by additional injected carriers. In this picture, the observed brightness enhancement does not necessarily arise from the suppression of non-radiative recombination channels, but rather from an increased optical response due to multiple, spectrally similar photon-emitting centers. As a consequence, the multi-photon suppression is reduced, potentially limiting the quantum emitter performance. As observed herein, when comparing the PLQY and anti-bunching data from the $\text{WSe}_2/\text{SiO}_2$, $\text{WSe}_2/\text{thin-Clinochlore}$ and $\text{WSe}_2/\text{thick-Clinochlore}$, we confirm the interplay between brightness enhancement and single-photon purity, governed by substrate-induced carrier injection and interface disorder. This understating can be substantial to achieve high-performance TMDC-SPEs by comprehending the interface mechanisms behind the conception of layered materials quantum devices.  
 
\subsection*{Conclusion}

In summary, our experimental results and theoretical description provide a robust investigation towards revealing a microscopic interpretation for the interface-driven mechanisms in a 2D-based non-classical light. Throughout a detailed study combining magneto-optical and quantum optical characterization, and scanning-probe measurements, we gained substantial knowledge of quantum emitters in a WSe$_2$ monolayer transferred onto a high-doped dielectric Clinochlore substrate. We reveal a pronounced interface-driven coupling that enhances the photoluminescence quantum yield by up to fivefold and substantially reshapes carrier dynamics through dielectric modulation of Coulomb interactions. Second-order autocorrelation measurements show a marked degradation of single-photon purity when the dielectric environment is modified, directly evidencing the active role of the interface, not yet explored in the literature of van der Waals based SPEs. Overall, our findings establish interface-induced dielectric modulation as a key design parameter for engineering high-quality single-photon emitters in van der Waals heterostructures, shifting the environment from a secondary consideration to a central element in quantum device optimization.

\section*{Methods}

\paragraph*{Sample preparation.} 
The mechanical exfoliation\cite{Novoselov2004} and the dry-transfer method\cite{CastellanosGomez2014} were performed through all 2D materials (WSe$_2$ crystal, Clinochlore and hexagonal boron nitride~(hBN)) used to fabricate our sample. The MoSe$_2$ MLs were obtained by exfoliating TMD bulk crystals synthesized through flux zone growth \cite{Nagao2017} and identified by optical contrast and $\mu$PL spectroscopy. To stack the van der Waals heterostructure, we used the polydimethylsiloxane (PDMS) stamp. 

\paragraph*{Optical, magneto-optical and quantum spectroscopy characterization.} 
The sample is mounted onto a piezo-stack for x–y–z nanopositioning in a closed-cycle cryostat reaching a minimal temperature of 3.9 K. A confocal microscope with a numerical aperture of 0.81 is placed within the cryostat for excitation and detection. Optical excitation is performed using the following lasers: (a) dual mode 660 nm diode-laser (continuous wave (CW) and 20 ps pulses) (b) 2 ps pulses from a doped fiber laser that are converted by an optical parametric oscillator (OPO). The PL emission is detected using a spectrometer equipped with thermo-electrically cooled charged-coupled device (CCD) with spectral resolution of 0.12~nm). For the TRPL experiment and the second-order autocorrelation measurement (using a Hanbury Brown and Twiss setup), the emission of the emitter is spectrally filtered using a set of angle-tuned thin-film  filters with bandwidth of 5~nm. Both measurements are performed using a superconducting nanowire single photon detectors, with a time resolution of $\sim50~$ps per channel, where the emission is sent to the detector through a $50:50$ fiber beam splitter. 
To perform the magneto-PL experiments, the sample was mounted on an x-y-z piezoelectric Attocube stage in a closed-cycle cryostat equipped with superconducting magnet coils under magnetic fields up to 9 T (Attocube - Attodry 1000). The measurements were collected at a temperature of 3.6 K using a continuous-wave (cw) linearly polarized laser with a photon energy of 1.88 eV (660nm). The photoluminescence (PL) signal was collimated using an aspheric lens (Attocube LT-IWDO, NA = 0.68), and the selection of circular polarization components was performed using appropriate polarizer and quarter wave plate. The PL signal was then dispersed by a 75 cm Andor spectrometer with gratings of 150 l/mm or 600 l/mm and detected by a Si CCD detector (Andor, Shamrock/iDus).

\paragraph*{Kelvin Probe Force Microscopy measurement.}

KPFM experiments were performed at in-situ growth laboratory (LCIS) of the Brazilian Nanotechnology National Laboratory (LNNano, Campinas/Brazil) using a Nanosurf FlexAFM microscope under N2 atmosphere with $<2\%$ of relative humidity. Pt-Cr metallic coated tips were used in tapping mode excited with 500 mV of free oscillation amplitude and $50\%$ of setpoint. The modulation electrical signal was set to 5 V of amplitude and 17 kHz of frequency. The sample substrate was grounded using conductive silver ink.

\section*{Supporting Information}
 
Additional PL data, second-order autocorrelation data fitting and the model for the Time Resolved Photoluminescence.

\section*{Conflict of interest}

The authors declare no competing interest.

\section*{Data availability statement}

The data that support the findings of this study are available from the corresponding author upon reasonable request. 

\section*{Acknowledgment}

The authors thank Prof. M. A. Fonseca (Federal University of Ouro Preto, UFOP) for providing the clinochlore mineral samples. The Kelvin probe force microscopy measurements were carried out at the in-situ growth laboratory (LCIS) of the Brazilian Nanotechnology National Laboratory (LNNano, Proposal No. 20250150), a facility of the Brazilian Center for Research in Energy and Materials (CNPEM), a private non-profit organization under the supervision of the Brazilian Ministry of Science, Technology, and Innovation (MCTI). The authors also acknowledge the Brazilian Synchrotron Light Laboratory (LNLS) for access to the Microscopic Samples Laboratory (LAM) under proposal LAM-2D: 20242176. This work was supported by the Fundação de Amparo à Pesquisa do Estado de São Paulo (FAPESP, Grants Nos. 22/10340-2, 22/08329-0, 23/08276-7, 23/13081-0, 24/05925-7, 24/00989-7, 25/09382-0 and 408783/2024-9). C.S. acknowledges support from the Coordenação de Aperfeiçoamento de Pessoal de Nível Superior (CAPES). Y.G.G. and I.B. acknowledge support from the Conselho Nacional de Desenvolvimento Científico e Tecnológico (CNPq, Research Fellowship Nos. 306971/2023-2 and  408521/2025-2). A. Koulas-Simos, C. C. Palekar, and S. Reitzenstein acknowledge financial support from the Deutsche Forschungsgemeinschaft (DFG) within the Priority Program SPP 2244 “2DMP”, Project Re2974/26-1 (ID 443416027), and from the Berlin Senate through Berlin Quantum.


\bibliographystyle{unsrt} 
\bibliography{reference}

\end{document}